\documentclass[12pt,preprint]{aastex}

\usepackage{subfigure}
\usepackage{longtable}
\usepackage{graphicx}
\usepackage{graphics}
\usepackage{keyval}
\usepackage{trig}

\begin{document}
\title{Catastrophic Photo-z Errors and the Dark Energy Parameter Estimates with Cosmic Shear}
\author{Lei Sun $^{1,2}$, Zu-Hui Fan$^{1}$, Charling Tao$^{2}$,
Jean-Paul Kneib$^{3}$, St\'ephanie Jouvel$^{3}$, Andr\'e
Tilquin$^{2}$}
\altaffiltext{1}{Department of Astronomy, Peking
University, Beijing 100871, China }
\altaffiltext{2}{Centre de
Physique des Particules de Marseille, CNRS/IN2P3-Luminy and
Universit\'e de la M\'editerran\'ee, Case 907, F-13288 Marseille
Cedex 9, France}
\altaffiltext{3}{Laboratoire d'Astrophysique
de Marseille - UMR 6110 - CNRS-Universit\'e de Provence-38 rue Fr\'ederic Joliot-Curie, 13013 Marseille,
France}
\email{sunl@bac.pku.edu.cn}

\begin{abstract}
We study the impact of catastrophic errors occurring in the
photometric redshifts of galaxies on cosmological parameter
estimates with cosmic shear tomography. We consider a fiducial
survey with 9-filter set and perform photo-z measurement
simulations. It is found that a fraction of $1\%$ galaxies at
$z_{\rm spec}\sim0.4$ is misidentified to be at $z_{\rm
phot}\sim3.5$. We then employ both $\chi^2$ fitting method and the
extension of Fisher matrix formalism to evaluate the bias on
the equation of state parameters of dark energy, $w_0$ and $w_a$, induced by
those catastrophic outliers. By comparing the results from both
methods, we verify that the estimation of $w_0$ and $w_a$ from
the fiducial 5-bin tomographic analyses can be significantly biased. To minimize the impact of this bias, two strategies can be followed: (A)  
the cosmic shear analysis is restricted to $0.5<z<2.5$
where catastrophic redshift errors are expected to be insignificant; (B)  
a spectroscopic survey is conducted for galaxies with $3<z_{\rm phot}<4$. We find  
that the number of spectroscopic redshifts needed scales as $N_{\rm spec} 
\propto f_{\rm cata}\times A$ where $f_{\rm cata}=1\%$ is the fraction of catastrophic redshift errors
(assuming a 9-filter photometric survey) and A is the survey area. For  
$A=1000 \hbox{ deg}^2$, we find that $N_{\rm spec}>320$ and 860 respectively in  
order to reduce the joint bias in $(w_0,w_a)$ to be smaller than $2\sigma$
and $1\sigma$. This spectroscopic survey (option B) will improve the Figure of Merit  
of option A by a factor $\times 1.5$ thus making such a survey strongly  
desirable.
\end{abstract}


\section{Introduction}

The weak gravitational lensing effect by large scale structures,
namely the cosmic shear, is believed to be a powerful tool to
measure the cosmological parameters including the equation of state
of dark energy. Over the last ten years, weak lensing observations
have already provided increasingly better constraints on the matter
density $\Omega_M$ and the amplitude of mass fluctuations often
represented by $\sigma_8$, the $rms$ amplitude of the smoothed and
linearly extrapolated mass density perturbations on the scale of
$8\hbox{ Mpc}\ h^{-1}$ (Hoekstra, Yee \& Gladders 2002, Jarvis et
al. 2003, Rhodes et al. 2004, Heymans et al. 2004, Hoekstra et al.
2006, Semboloni et al. 2006,  Massey et al. 2007, Benjamin et al.
2007, Fu et al. 2008). Their contributions to the constraints on the
total neutrino mass have also been investigated
(Tereno et al. 2008, Gong et al. 2008, Ichiki et al. 2008, Li et al. 2008).
Several more ambitious weak lensing surveys
targeting at hundred to thousand times larger areas, such as
DES\footnote{http://www.darkenergysurvey.org},
LSST\footnote{http://www.lsst.org},
JDEM\footnote{http://jdem.gsfc.nasa.gov} and
EUCLID\footnote{http://sci.esa.int/science-e/www/object/index.cfm?fobjectid=42266},
are being planned aiming at exploring the properties of dark
energy in great detail.

Future large surveys are expected to be able to control statistical errors to
insignificant levels. Therefore different systematics can become
dominant sources of contaminations in future weak lensing studies, and
should be investigated carefully. These include the effect of the
point spread function (PSF) (e.g., Amara \& Refregier 2008), the
methodology in the shear measurement (Bacon et al. 2001, Erben et al.
2001, Hirata \& Seljak 2003, Heymans et al. 2006, Massey et al. 2007, Miller et al. 2007,  Kitching et al. 2008, Bridle et al. 2008, Ngan et al. 2008, Semboloni et al.
 2008), redshift errors (Ma et al. 2006,
Huterer et al. 2006), intrinsic alignments of source galaxies
(see Heavens 2001 for a review, Bridle \& King 2007, Fan 2007),
uncertainties in the computation of the non-linear power spectrum
(see discussion in Huterer 2002, Van Waerbeke et al. 2002), and so
on. In this paper, we focus on the impact of the photometric redshift
errors on cosmological studies.

The necessity of measuring vast amount of galaxy redshifts in cosmic shear observations
makes the multi-color photometric determination of redshifts the only feasible
way in obtaining the redshift information of source galaxies.
However the inaccuracy of photometric redshifts can contaminate considerably
the cosmological information embedded in cosmic shears (Huterer et
al. 2006; Ma et al. 2006; Amara \& Refregier 2007; Wittman et al. 2007; Abdalla et al.
2008; Margoniner \& Wittman 2008; Stabenau et
al. 2008). Taking into account the bias and the scatter of
$z_{\rm phot}$ with respect to $z_{\rm spec}$, Ma, Hu \& Huterer (2006)
analyze the corresponding degradation of the constraints on dark energy.
It is found that these errors can completely erase the additional information from tomographic lensing
observations. In order to make full use of the advantage of the
tomographic binning, both the bias and scatter of $z_{\rm phot}$ have to
be known to better than 0.003-0.01. Amara \& Refregier (2007) and
Abdalla et al. (2008) pay special attention to the so-called
catastrophic errors in $z_{\rm phot}$, the outliers with
$|z_{\rm spec}-z_{\rm phot}|>1$, and explore their effects on the figure of
merit of dark energy parameter constraints.

While most of the existing studies concentrate on the error range of
parameter constraints, the systematic bias in the determination of
cosmological parameters due to the photometric redshift errors
should also be carefully analyzed. A large systematic bias can lead
to a completely wrong conclusion on the values of the cosmological
parameters although the statistical errors may be small (Amara \& Refregier 2008).
Our study presented in this paper emphasizes precisely this aspect
of effects on dark energy parameter estimates arising from
catastrophic failures in photo-z measurements. We focus on a
survey with similar design as described in Aldering et al. (2004), with its
9-filter set described also in Dahlen et al. (2008) and Jouvel et al. (2009 in
prep.). We analyze in particular the bias on dark energy parameter estimates
induced by the catastrophic failures occurred around $(z_{\rm spec}, z_{\rm phot})\sim (0.4, 3.5)$.
And we investigate in detail the requirements of the number of spectroscopic redshift
measurements to retrieve the correct estimates.

The outline of the paper is as follows. In \S 2, we introduce the
methods to describe the lensing observables and to evaluate the
biases. We present the results of bias in \S 3. In \S 4, we discuss
the calibration requirements to remove the biases due to the
catastrophic failures. Discussions and conclusions are given in \S
5.

\section{Methodology}
\label{sect:meth}

In this section, we describe the formalism of lensing observables.
Based on simulation results, we then model the source galaxy
distribution with catastrophic failures taken into account. Finally
we discuss the methods to evaluate the corresponding biases on dark
energy parameter estimates, without and with spectroscopic calibration.

\subsection{Lensing Observables}
\label{sect:lens}

The convergence power spectrum $P_{ij}^{\kappa}(\ell)$ for the $i$th and $j$th redshift bin can be
expressed as (see Kaiser 1992, Kaiser 1998, Ma et al. 2006)
\begin{equation}
P_{ij}^{\kappa}(\ell) = \int_0^{\infty} dz \,{W_i(z)\,W_j(z)}{ H(z) \over D^2(z)}\,
 P\! \left (k_{\ell},  z\right ),
\label{eq:pk_l}
\end{equation}
\noindent where $H(z)$ is the Hubble parameter and $D(z)$ is the comoving angular diameter distance. The matter power spectrum $P(k_{\ell}, z)$ with
$k_{\ell} = \ell /D(z)$ is computed from the linear transfer
function of Bardeen et al. (1986), and then the non-linear fitting
function of Peacock \& Dodds (1996). The weighting function $W_i$ is
given by
\begin{eqnarray}
 W_i(z) &=& {3\over 2}\,\Omega_m\, {H_0^2 D(z) \over H(z)}(1+z)\nonumber  \\
 &&\times
 \int_z^\infty
 dz^{\prime} {n_i(z^{\prime})}
{D_{LS}(z,z') \over D(z')} \,,
\label{eqn:weights}
 \end{eqnarray}
where $D_{LS}(z,z')$ is the angular diameter distance between $z$
and $z'$, and $n_i$ is the galaxy distribution (normalized to unity)
of the $i$th redshift bin. We employ the overall galaxy distribution
of the form (Smail et al. 1994)
\begin{equation}
n(z) \propto z^{\alpha} \exp\left [-(1.41z/z_{\rm med})^\beta\right ]
\label{eq:nz}
\end{equation}
where $z_{\rm med}$ is the median redshift of a survey and we take $\alpha=2,\ \beta=1.5 $.

Considering only the shot noise and the Gaussian sample variance,
the covariance matrix of the lensing observables can be written as (e.g., Huterer et al. 2006, Ma et al. 2006)
\begin{equation}
{\rm Cov}\left [C^{\kappa}_{ij}(\ell'), C^{\kappa}_{kl}(\ell)\right ] =
{\delta_{\ell \ell'}\over (2\ell+1)\,f_{\rm sky}\,\Delta \ell}\,
\left [ C^{\kappa}_{ik}(\ell) C^{\kappa}_{jl}(\ell) +
  C^{\kappa}_{il}(\ell) C^{\kappa}_{jk}(\ell)\right ].
\label{eq:cov}
\end{equation}
\noindent where $\Delta \ell$ is the band width in multipole $\ell$, and $f_{\rm
sky}$ is the fractional sky coverage of a survey.
The total power spectrum $C^{\kappa}_{ij}$ is given by
\begin{equation}
C^{\kappa}_{ij}=P_{ij}^{\kappa} +
\delta_{ij} {  \sigma_{\gamma}^2  \over {n}_g^i} \,,
\label{eq:c_obs}
\end{equation}
\noindent where $\sigma_{\gamma}^2$ is the variance of each
component of the intrinsic ellipticity of source galaxies, and
$n_g^i$ is the surface number density of galaxies in the
$i$th bin.

In our study, we consider a fiducial space-based survey
with $f_{\rm sky}=0.025$, corresponding to 1000 $\hbox{deg}^2$,
the summation of $n_g^i$, i.e., the total surface number density of
source galaxies $n_g^{\rm tot}$, is taken to be $100$ arcmin$^{-2}$,
$\sigma_{\gamma}$=0.22 and $z_{\rm med}$=1.26. We are aware of the
likely extension of the survey area. Thus in analyzing the requirement of the
spectroscopic redshift calibration, we derive a scaling relation that allows us
to estimate the corresponding calibration requirement for any survey area and
any catastrophic error fraction.

Investigations by Ma et al. (2006) show that tomographic analyses
with multi-bins of source galaxies (denoted as $N_{\rm bin}$)
can improve the statistical errors on cosmological
parameter constraints considerably in comparison with those of 2-D analyses.
The improvement increases with the increase of $N_{\rm bin}$.
On the other hand, we cannot gain much further improvement with $N_{\rm bin}>5$. Hence
throughout the paper, we consider lensing tomography with $5$ equally spaced
redshift bins within the range of $z=[0,\ 3]$. All galaxies with $z>
3$ are simply added to the last bin. The resulting redshift
binning is illustrated in Figure \ref{Fig:nz}.

\subsection{Modeling galaxy distribution with catastrophic failures}
\label{sect:model}

Photometric estimations of galaxy redshifts through multi-waveband
observations largely depend on the characteristic features of the
spectral energy distribution (SED) of galaxies, such as the Lyman
break and Balmer/$4000 \AA$ break. The locations of the features
give us the redshift information of the galaxies. Therefore the
accuracy of photometric redshifts is sensitive to the wavelength
coverage of observations. It is also affected by the
observational signal-to-noise ratio and the estimation method.
Besides, the spectroscopic calibration plays an important
role in improving the accuracy of photo-z, which shall be discussed
specifically in \S 4.

To realistically assess the accuracy of photometric redshifts for
our survey, we first consider a 9-filter set which covers
a wavelength range from visible blue to infrared with $\lambda \sim
390\hbox{ nm}$ to $\sim 1700\hbox{ nm}$ (For specific definition, see Dahlen et al. 2008). Then a series of
simulations are carried out . The details of the simulations will be
presented in Jouvel et al. (2009 in prep.). Here we give a
brief description. The simulations are performed with the 'Le Phare'
software. First we generate galaxies according to the GOODS
luminosity function by Dahlen et al. (2005), each with an assigned
spectrum based on the Coleman Extended templates (Coleman et al.
1980). This set of templates contains $66$ spectra obtained by
interpolating between $5$ standard spectra: elliptical, Sbc, Scd,
irregular and star forming galaxies. We then apply the filtering
process to the simulated spectra with the 9-filter set.
We thus obtain the SED for each simulated galaxy.
The magnitudes of galaxies are randomly assigned in accordance with
the properties of our fiducial survey. The photometric redshift for each
galaxy is then computed using the LePhare photometric redshift
code. The results are shown in Figure \ref{Fig:snap}.

It is seen that scatters and biases around $z_{\rm spec}$ exist for
$z_{\rm phot}$. Besides, there are notable islands in the $z_{\rm
spec}$-$z_{\rm phot}$ plot. The dominant one is located at $z_{\rm
spec}\sim 0.4$ and $z_{\rm phot}\sim 3.5$, which accounts for more
than $50\%$ of the galaxies with $|z_{\rm phot}-z_{\rm spec}|> 1$.
This island obviously comes from the confusion of the $4000 \AA$
break with the Lyman break in galaxy SEDs due to the limited
wavelength coverage. It is found that the fraction of galaxies with
such large catastrophic errors is about $1 \%$ of the total number
of galaxies. In our study, we particularly analyze the effect of
this island on the constraints of dark energy parameters.

To characterize the true distribution of galaxies whose photometric
redshifts suffer from catastrophic failures, we use the following
Gaussian distribution with the form
\begin{equation}
A_{\rm cata}\exp\bigg [-{(z-z_m)^2\over 2\sigma^2}\bigg ],
\label{eq:cata}
\end{equation}
where the 'true' mean redshift of the island $z_m=0.398$ and the
standard deviation $\sigma=0.108$ are obtained from fitting to the
spectral(true)-z distribution histogram of the island. The parameter
$A_{\rm cata}$ is determined by normalizing the Gaussian
distribution to the overall fraction of catastrophic failures
$f_{\rm cata}=0.01$. Then in a tomographic division according to
galaxies' photometric redshifts, this part will be mistakenly
shifted from low-z bin to another high-z bin. Specifically, for the
$N_{\rm bin}=5$ tomography we consider, the $1 \%$ galaxies with
their true-z falling in the first bin are assigned to the last (5th)
bin whose redshift range is $z_{\rm phot} > 2.4$.

\subsection{Evaluation of biases on dark energy parameter estimates}
\label{sect:evaluation}

Once the distribution of catastrophic failure fraction is known, we
can evaluate the biases it induces on dark energy parameter
estimates.

As mentioned in the previous subsection, we need to be aware of the
fact that observationally, source galaxies are assigned to different
redshift bins in terms of their 'observed' photometric redshifts
rather than their true redshifts. Hence, a fraction of low redshift
galaxies can be mistakenly distributed to high-redshift bins due to
the catastrophic errors. As a consequence, weak lensing signals from
those relevant tomographic bins are perturbed and biases on
cosmological parameter estimates may arise, if models used to fit the data were not
modified correspondingly.

Concerning the fiducial survey in this study, we know from
simulations that there will be $f_{\rm cata}=0.01$ galaxies being
misplaced into the 5th bin (denoted as 'Bin-5'), whilst their true
redshifts are within the range of the first bin (denoted as
'Bin-1'). The true distribution of this catastrophic failure part
can be described with Eq. (\ref{eq:cata}) and will be labeled as
Bin-cata for convenience in the following discussion. Note that the
true z-range of Bin-cata is within the z-range of Bin-1.

For Bin-1, when Bin-cata galaxies within it are misplaced in Bin-5 due
to catastrophic redshift errors, the number of galaxies contained in
Bin-1 is reduced. The redshift distribution of the galaxies left in
it also changes somewhat in comparison with the true distribution
without the catastrophic errors. The number change in Bin-1 does not
affect the lensing signal because the galaxy distribution within the
bin is always normalized to unity. The shape change of the redshift
distribution in the bin may have some effects. On the other hand,
however, this change may already be included in the modeled overall
redshift distribution of galaxies derived from photometric
redshifts. Thus in our consideration, we assume that the
misplacement of Bin-cata does not affect the lensing signal from Bin-1.

For bins in between Bin-1 and Bin-5, obviously the lensing signals
should be kept unperturbed. Combining the previous analysis for Bin-1,
we thus get $\bar{C}_{ij}^{\kappa}(\ell) = C_{ij}^{\kappa}(\ell)$, where $\rm i,j=1,...,4$.

To the contrary, the lensing signals from Bin-5 get perturbed. As
shown in the bottom-left panel of Figure \ref{Fig:model}, it in fact
consists of two parts of galaxies, one with their true redshifts
falling in Bin-5 and the other being a contamination from Bin-cata.
Correspondingly, besides the auto spectra of Bin-5, those cross
spectra with one of the bins being Bin-5 will also be affected. In
this case, those convergence power spectra can be expressed as
\begin{equation}
P_{i5}^{\kappa}(\ell) = \int_0^{\infty} dz \,{W_i(z)\,W_5(z)}{ H(z) \over D^2(z)}\,
 P\! \left (k_{\ell},  z\right ),\ \ \ \ \  (\rm i=1,...,5)
\label{eq:pk_i5}
\end{equation}
\noindent with
\begin{equation}
 W_5(z) = \cdot\cdot\cdot \, \int_z^\infty
 dz^{\prime} {n_5(z^{\prime},f_{\rm cata}, z_m, \sigma)}
{D_{LS}(z,z') \over D(z')} \,
\label{eq:w5}
 \end{equation}
\noindent where we omit the antecedent terms identical to those in
Eq. (\ref{eqn:weights}). The true (spectral) galaxy distribution of
Bin-5 can be written as
\begin{eqnarray}
n_5(z,f_{\rm cata}, z_m, \sigma) &=& {1 \over f_5+f_{\rm cata}}\, \left [ n_5(z)+n_{\rm cata}(z,f_{\rm cata}, z_m, \sigma) \right ]\nonumber \\
&=&{1 \over f_5+f_{\rm cata}}\, \{ n_5(z)+A_{\rm cata}\exp\bigg [-{(z-z_m)^2\over 2\sigma^2}\bigg ] \}
\label{eqn:n5}
\end{eqnarray}
\noindent where $1/(f_5+f_{\rm cata})$ is to normalize $n_5(z,f_{\rm
cata}, z_m, \sigma)$ to be unity, with $f_5$ being the overall
fraction of galaxies with their true redshifts falling in Bin-5.
Here the right side terms $n_5(z)$ and $n_{\rm cata}$ are normalized
to their overall fraction $f_5$ and $f_{\rm cata}$, respectively.
$A_{\rm cata}$ is thus determined by $f_{\rm cata}$, as mentioned in
the paragraph following Eq. (\ref{eq:cata}). With the substitution
$n_g^5 \rightarrow n_g^5+n_g^{\rm cata}$ in Eq. (\ref{eq:c_obs}),
finally we get the 'observed' total spectra $C_{i5}^{\kappa}(\ell,
f_{\rm cata}, z_m, \sigma)$, where $f_{\rm cata}=0.01, z_m=0.398,
\sigma=0.108$ and $i=1,...,5$.

In practical analysis, when calculating the `theoretically expected'
lensing signals for Bin-5, we have no knowledge, in the case without
spectroscopic calibration, of the fraction of low redshift galaxies
and regard all the galaxies being at high redshifts; or in the case
with inadequate spectroscopic calibration, we have inaccurate
knowledge therefore can make unprecise estimation of the fraction
and distribution of low-z galaxies (see Figure \ref{Fig:model}). The
former can be denoted as $\bar{C}_{i5}^{\kappa}(\ell, 0, c, c)$,
with c being an arbitrary constant. The latter case gives
$\bar{C}_{i5}^{\kappa}(\ell, \bar{f}_{\rm cata}, \bar{z}_m,
\bar{\sigma})$, where the estimations $\bar{f}_{\rm cata},
\bar{z}_m$ and $\bar{\sigma}$ can deviate from their true values
depending on the calibration size.

Consequently systematic biases may arise when the theoretically
expected signals $\bar{C}_{ij}^{\kappa}$ are compared with the
observed ones $C_{ij}^{\kappa}$ to constrain cosmological
parameters. As an approximation, a simple extension of the Fisher
matrix formalism can be used to compute the bias in cosmological
parameters with (see Huterer et al. 2006, Amara \& Refregier 2008)

\begin{equation}
\delta p_i  =  F_{ij}^{-1}\, \sum_{\ell, \alpha, \beta}
\left [C_{\alpha}^{\kappa}(\ell)-\bar{C}_\alpha^{\kappa}(\ell)\right ]
 {\rm Cov}^{-1}\left [ C_{\alpha}^{\kappa}(\ell),
   C_{\beta}^{\kappa}(\ell)\right ]\,
   {\partial \bar C_{\beta}^{\kappa}(\ell) \over \partial p_j}
\label{eq:bias}
\end{equation}

\noindent where $\alpha$ and $\beta$ each denote a pair of redshift
bins. The term $\left
[C_{\alpha}^{\kappa}(\ell)-\bar{C}_{\alpha}^{\kappa}(\ell)\right ]$
reflects the bias on the shear signals. The covariance of the cross
power spectra is given in Eq. (\ref{eq:cov}). $F_{ij}^{-1}$ is the
inverse of the Fisher matrix

\begin{equation}
F_{ij} = \sum_{\ell} \,
\left ({\partial {C}\over \partial p_i}\right )^T\,
{\rm Cov}^{-1}\,
{\partial { C}\over \partial p_j},\label{eq:fish}
\end{equation}
\noindent where ${ C}$ is the column matrix of the convergence power
spectra and ${\rm Cov}^{-1}$ is the inverse of the covariance
matrix.

Eq. (\ref{eq:bias}) is a linear approximation around the fiducial model.
On the other hand, the true extent of bias can be quantified by a
direct chi-square fitting method, by minimizing

\begin{equation}
\chi^2 =  \sum_{\ell, \alpha, \beta}
\left [C_{\alpha}^{\kappa}(\ell)-\bar{C}_\alpha^{\kappa}(\ell)\right ]
 {\rm Cov}^{-1}\left [ C_{\alpha}^{\kappa}(\ell),
   C_{\beta}^{\kappa}(\ell)\right ]\,
   {\left [C_{\beta}^{\kappa}(\ell)-\bar{C}_\beta^{\kappa}(\ell)\right ]}.
\label{eq:chi2}
\end{equation}

In our analyses, we will utilize both bias evaluation methods.
The comparison of their results permits us to quantify the extent to which
the linear approximation is valid. We shall come back to this in
the following section, associated with the specific study for our fiducial survey.

Throughout our study, a 7 parameter fiducial flat cosmological model
is assumed. For equation of state of dark energy, we consider the
form $w(z) = w_0 + {w_a}z/(1+z)$ (Chevallier \& Polarski 2001). The fiducial values
of the 7 parameters are: $\Omega_m = 0.27, w_0=-1, w_a=0, h=0.72,
\sigma_8 = 0.78, \Omega_b = 0.0446, n=0.96$, respectively.

All the weak lensing related calculations are done with the ICOSMO
package (Refregier, version 2005\footnote{for a latest version, see
http://www.icosmo.org/ and Refregier et al. (2008)}), with
modifications from 2D weak lensing to tomographic form for our
specific study. The $\chi^2$ fitting tool is
MPFIT\footnote{http://cow.physics.wisc.edu/$\sim$craigm/idl/fitting.html}
from C. Markwardt.

\section{The bias on dark energy parameter estimates}
\label{sect:result}

To demonstrate clearly their significance, in this section,
we present the results of bias on dark energy
parameters due to catastrophic failures if there is no
spectroscopic calibration for the problematic Bin-5.
We apply both methods by $\chi^2$ fitting and by computation with the extension of
Fisher matrix formalism. In the next section, we will focus on the requirements for the
spectroscopic calibrations.

First, we employ the $\chi^{2}$ fitting method. As described in the
previous section, for Bin-5, the observed
signal is a combination of the two parts illustrated in the bottom-left
panel of Figure \ref{Fig:model}. On the other hand, the model
calculation is done with the galaxy distribution illustrated in the
top panel of Figure \ref{Fig:model}. The $\chi^{2}$ fitting
is then applied to find the `best-fit' values of the cosmological
parameters by comparing the model-predicted lensing signals with the
`observational data'. In this fitting process, the `measurement
errors' for each `data point' are theoretically calculated using a
binned form of Eq. (\ref{eq:cov}).

In our calculations, the $\ell$ range is taken to be $[50, 3000]$.
We discard weak lensing information from all those scales beyond
$\ell=3000$ to avoid complications from various small-scale effects,
such as baryonic cooling (White 2004, Zhan \& Knox 2005, Jing et al.
2006) and non-Gaussianity (White \& Hu 2000, Cooray \& Hu 2001). We
apply binning procedures to $\ell$. According to Hu \& White 2001,
the bin width should be at least twice as big as $\ell_{\rm field}$,
in order to reduce the covariance between the bins arising from the
survey geometry. We then take $10$ bins, uniform in logarithmic
scale over the range $\ell=[50,3000]$.

In the right panel of Figure \ref{Fig:bias}, we show the resulting
bias on the equation of state of dark energy obtained from $\chi^2$
fitting. Gaussian priors $\sigma(\Omega_b)=0.01$ for $\Omega_b$ and
$\sigma(p_i)=0.05$ for others are applied upon the hidden parameters in
our 7-parameter fiducial model. These priors are consistent with the current constraints.
It is found that the $1\%$ catastrophic failure fraction can bring large biases on $w_0$
and $w_a$ in the tomographic 5 z-bins case, highly significant
compared to their statistical errors. The `best fit' values are $w_0=-3.5$ and
$w_a=3.8$ in comparison with the fiducial ones $w_0=-1$ and $w_a=0$.

The results from the extended Fisher matrix formalism are shown in
the right panel of Figure \ref{Fig:bias}. The same $\ell$-binning
process and Gaussian priors are applied. The `best fit' values
are found to be $w_0=-4.2$ and $w_a=6$. Thus the linear approximation results in a
notably larger bias than those of the $\chi^2$ fitting. This significant discrepancy
reveals clearly the necessity to go beyond the linear approximation from the Fisher
matrix formalism.

In summary, our analyses show that for our fiducial survey with
a 9-filter set, the catastrophic photo-z errors with $f_{\rm cata}\sim
1\%$ can bias the dark energy parameters at a level far exceeding the statistical
errors if no further spectroscopic calibration is conducted.

Our calculations are done for a survey area $A=1000\hbox{ deg}^2$,
corresponding to a sky coverage $f_{\rm sky}=0.025$. Increasing the survey area
reduces the statistical errors with $\sigma \propto 1/\sqrt{f_{\rm
sky}}$. On the other hand, the absolute biases on cosmological
parameters should not change much. This can be seen clearly from Eq.
(\ref{eq:bias}) in which $f_{\rm sky}$ enters through $F_{ij}^{-1}$
and $\rm Cov^{-1}$. Since both are proportional to $f_{\rm
sky}^{-1}$, their dependencies on $f_{\rm sky}$ cancel out. Thus
with a larger sky coverage, the effect of the bias relative to the
statistical errors would be even larger if the catastrophic fraction
$f_{\rm cata}$ remains  $\sim 1\%$.

\section{Requirements of Spectroscopic Calibrations}
\label{sect:cali}

The large bias seen in Figure \ref{Fig:bias} clearly reveals the necessity to
take extra measures to reduce the effects of the catastrophic
redshift errors. Among others, the spectroscopic calibration can
play an crucial role in this regard. It is known that the
spectroscopic calibration can be done either by taking spectra of a
sub-sample of source galaxies after photo-z measurements or by
employing a sophisticated technique where the spectral information
is included in the determination of photometric redshifts. We
consider the former case in our analyses, and explore the
requirements of the spectroscopic calibration. Specifically, we
study the needed number of galaxies with measured spectroscopic
redshifts in order to retrieve the unbiased parameter estimates.

\begin{figure}
\epsscale{0.5}
\plotone{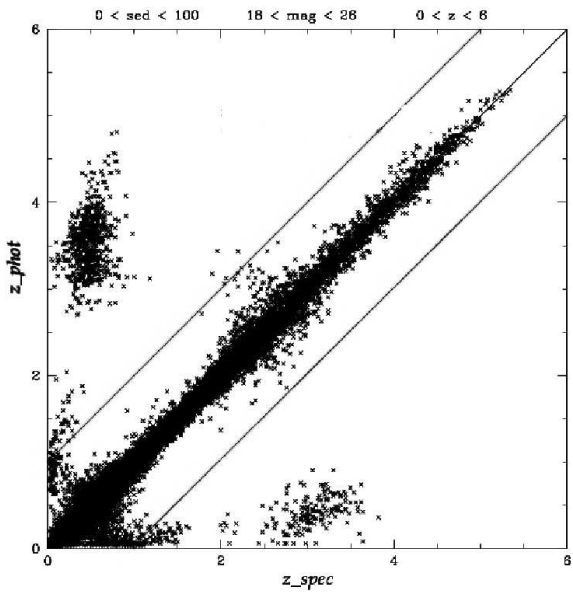}
\caption{Result of LePhare
simulation for our fiducial survey with a 9-filter set. A magnitude cut
of 26 AB is set to fit the median redshift of the survey. The $2$
grey lines show the borders inside which the relation $|z_{\rm
spec}-z_{\rm phot}|<1$ holds.}
\label{Fig:snap}
\end{figure}

\begin{figure}
\epsscale{0.5}
\plotone{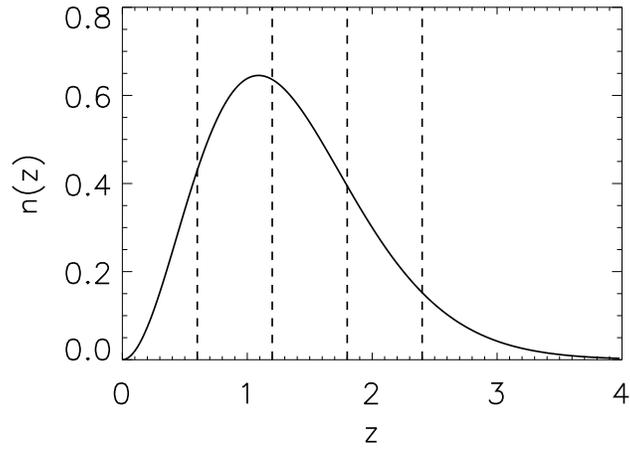}
\caption{Source galaxy distribution
$n(z)$. The solid curve is the overall galaxy distribution defined
in Equation (\ref{eq:nz}). 5 tomographic bins are equally spaced
between $0<z<3$, denoted by dashed vertical lines. All galaxies with
$z> 3$ are simply added into the last bin.}
\label{Fig:nz}
\end{figure}

\begin{figure}
\epsscale{1}
\plotone{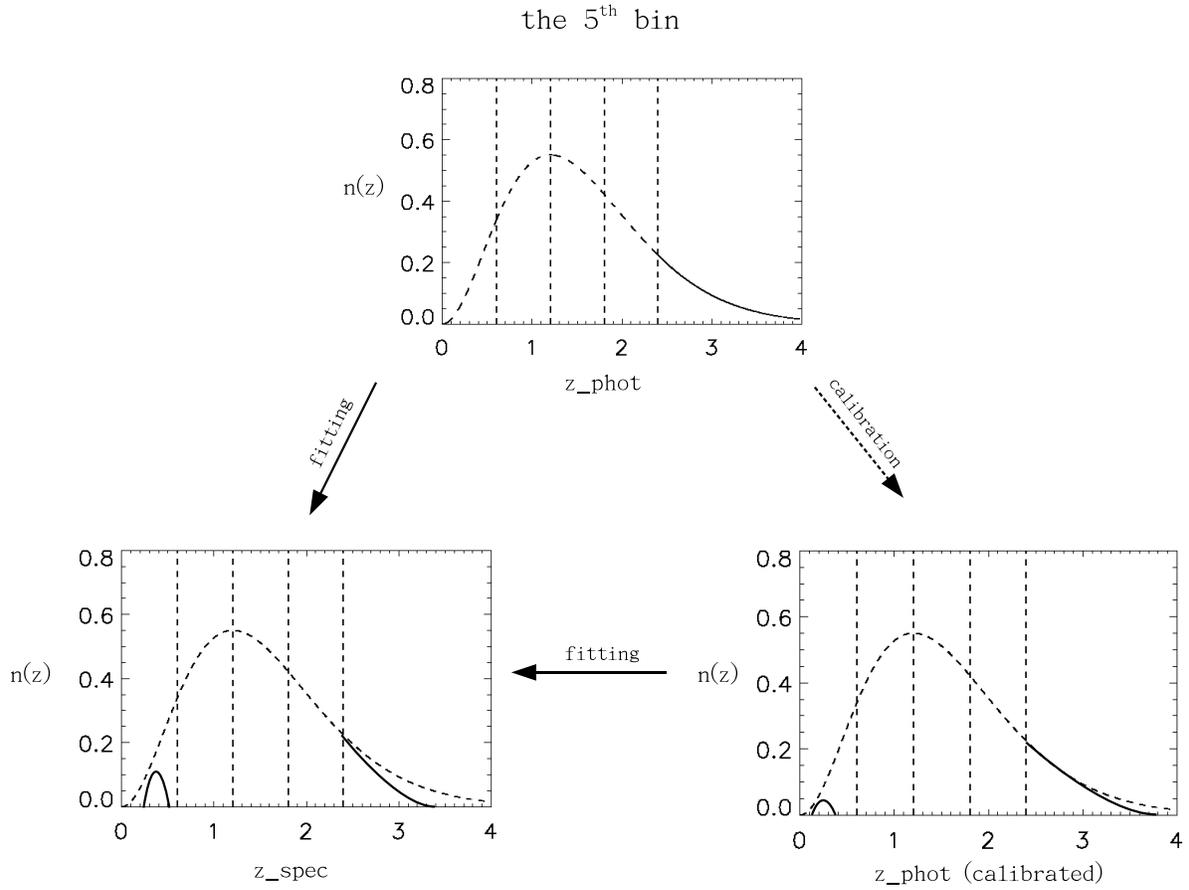}
\caption{\small A sketch to explain where the bias comes from in cases without and with calibration. The top panel shows the 'observed' photo-z distribution of Bin-5.  Bottom-left: the true (spectral) distribution of Bin-5.
Bottom-right: an example of the calibrated distribution of Bin-5, derived from
an individual realization of calibration simulations.}
\label{Fig:model}
\end{figure}

\begin{figure}
\epsscale{1}
\centering
\plottwo{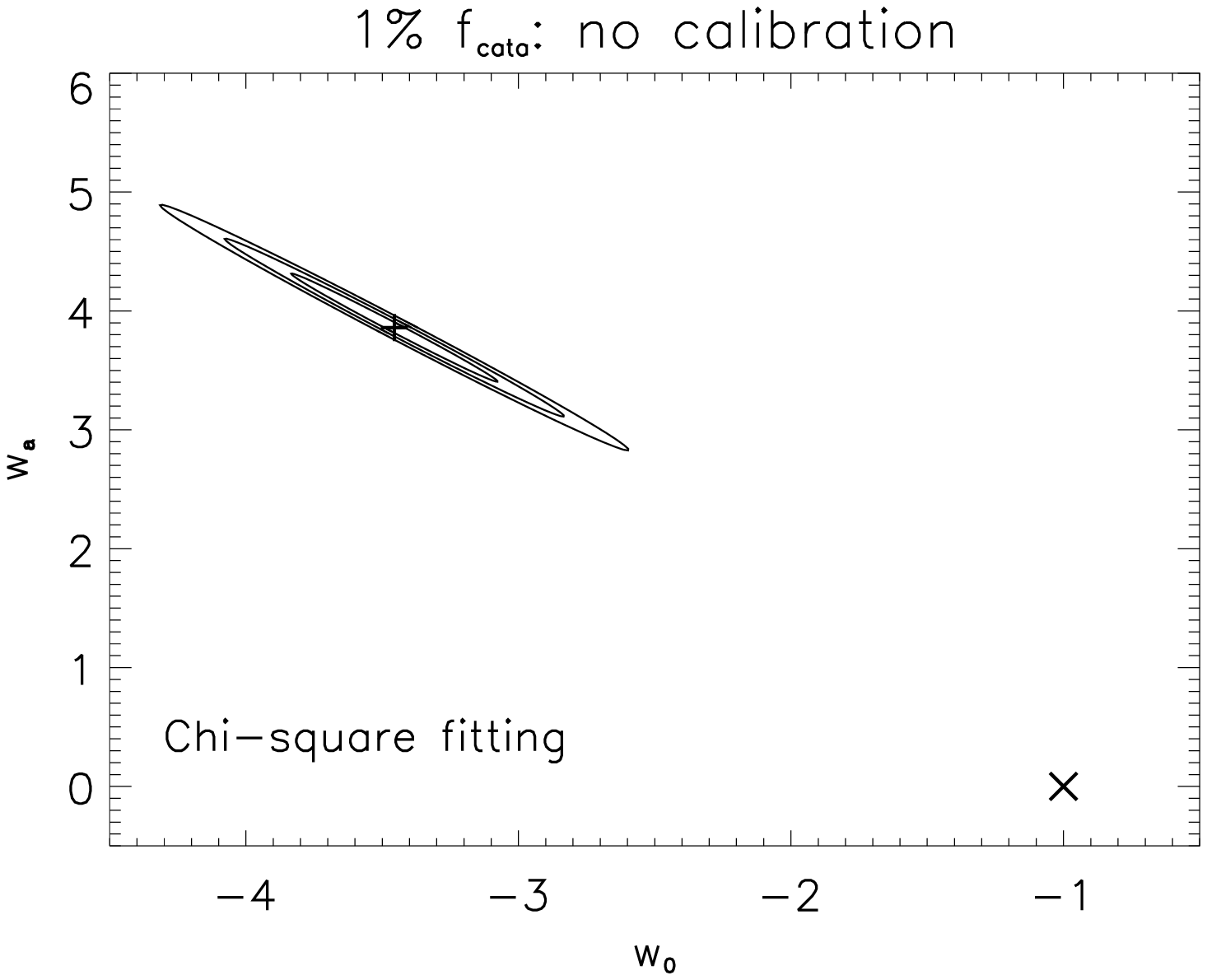} {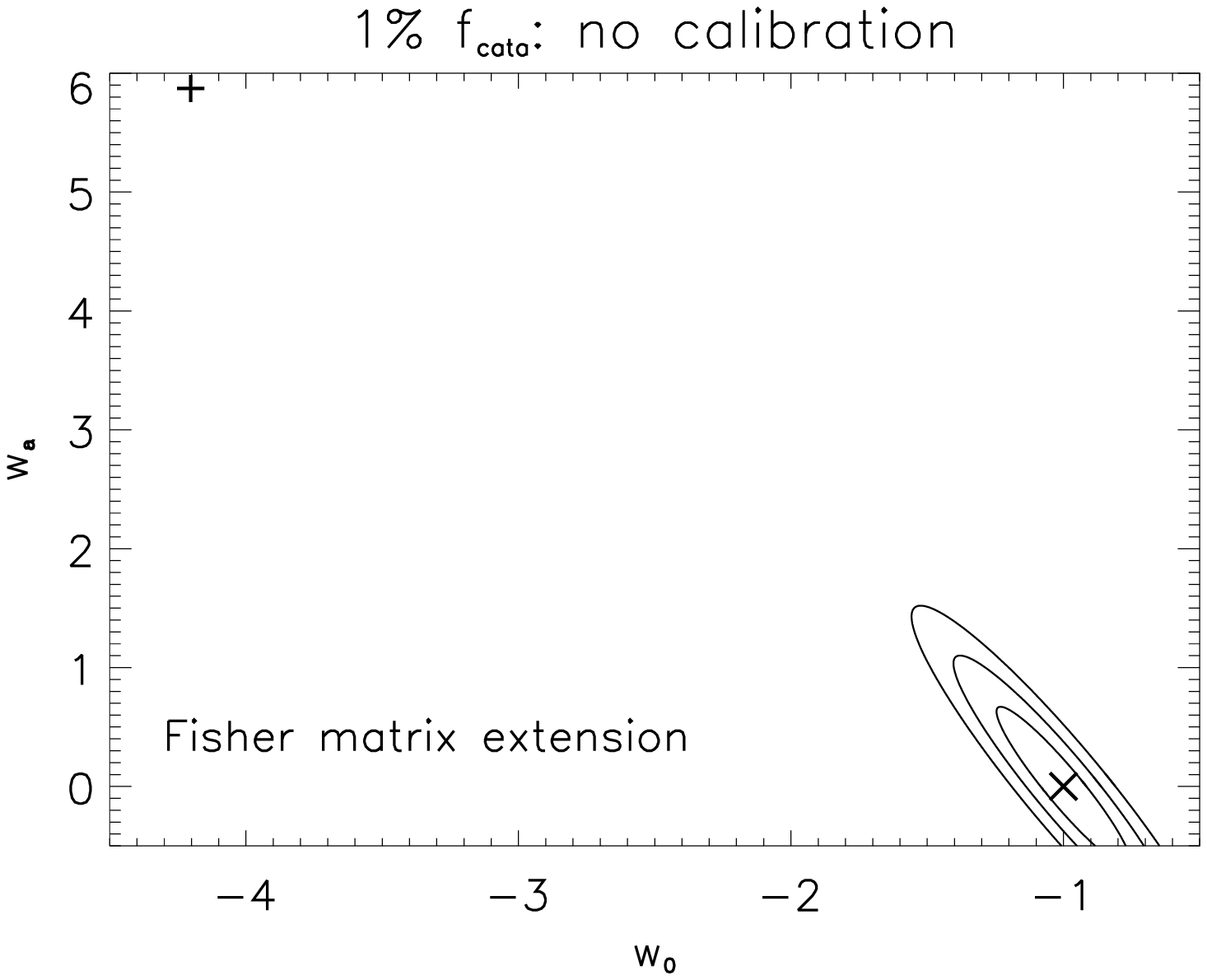}
\caption{\small  Biases on the dark energy parameters $w_0$ and $w_a$ in 5
z-bins tomography, from $\chi^2$ fitting (left panel) and from Fisher
matrix calculations (right panel), respectively. The fiducial value is denoted by
the oblique cross, and the plus represents the 'best fit' value.
The error contours ($1,2,3\sigma$, inside-out) are given at the
best fit value for the fit result and at the
fiducial value for Fisher matrix calculations, respectively.
Gaussian priors $\sigma(\Omega_b)=0.01$ for $\Omega_b$ and
$\sigma(p_i)=0.05$ are applied upon all other hidden parameters.}
\label{Fig:bias}
\end{figure}

Since we particularly analyze the `island' catastrophic failures
around $(z_{\rm spec}, z_{\rm phot})\sim (0.4,3.5)$ seen in Figure \ref{Fig:snap},
we consider the calibration for galaxies with their photo-z in the range $3<z_{\rm phot}<4$.
We assume that our simulation results shown in Figure \ref{Fig:snap} reflect
the real relation between photo-z and the true redshifts for observed galaxies.
In other words, we take the simulation results as the fiducial distribution with
an overall $f_{\rm cata}=1\%$. We then investigate how many spectra are needed
for the sub-sample of galaxies with $3<z_{\rm phot}<4$ in order to gain some
information on the catastrophic contamination, mainly the value of $f_{\rm cata}$.

Our analyzing procedures are as follows. We use the simulated
galaxies with $3\le z_{\rm phot}\le 4$ as our parent fiducial
sample. Then for a given $N_{\rm spec}$, where $N_{\rm spec}$ is the
number of galaxies chosen for spectral measurements, we perform a
large number of Monte Carlo random samplings of $N_{\rm spec}$
galaxies from the parent catalog. For each realization, we first
compute the estimated fraction of catastrophic failures
$\bar{f}_{\rm cata}$ by counting the number of galaxies among
$N_{\rm spec}$ that are identified in the realization to be in
Bin-cata, taking into account the overall fraction of galaxies that
have $3\le z_{\rm phot}\le 4$ . With this sub-sample of Bin-cata
galaxies, we then fit their distribution to the form given by Eq.
(\ref{eq:cata}) to obtain the estimated $z_m$ and $\sigma$, denoted
as $\bar{z}_m$ and $\bar{\sigma}$. With these calibrated values of
$\bar{f}_{\rm cata}$, $\bar{z}_m$ and $\bar{\sigma}$, we are able to
calculate the model-predicted signals. By comparing them with those
from the fiducial distribution, we therefore can evaluate the biases
on cosmological parameters for each individual realization. The
biases arise from the differences between the calibrated
($\bar{f}_{\rm cata}$, $\bar{z}_m$, $\bar{\sigma}$) and the fiducial
($f_{\rm cata}$, ${z}_m$, ${\sigma}$). With the results from all the
realizations, we can further evaluate statistically the level of
bias from $N_{\rm spec}$ calibrations.

\begin{figure}
\epsscale{1}
\centering
\plottwo{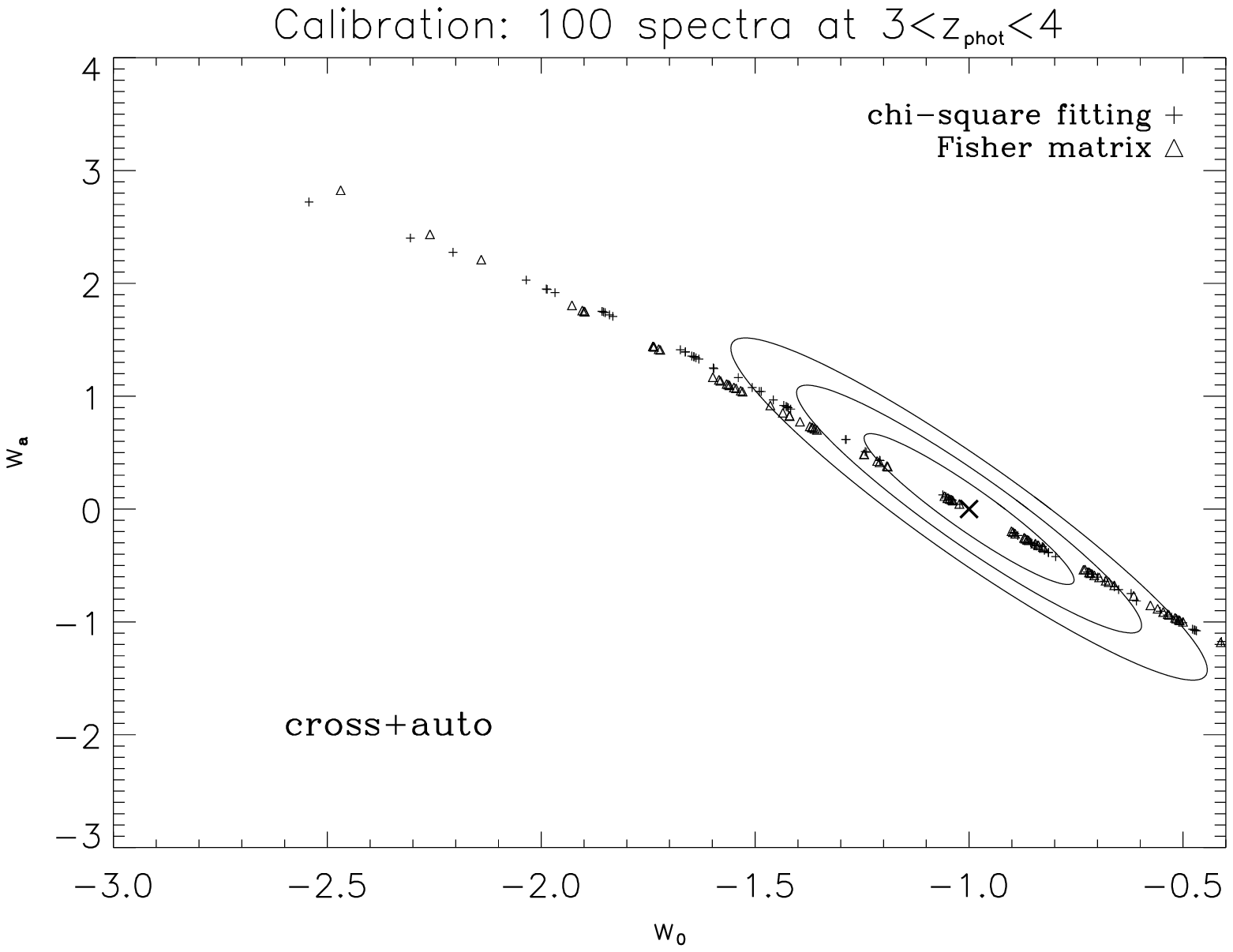} {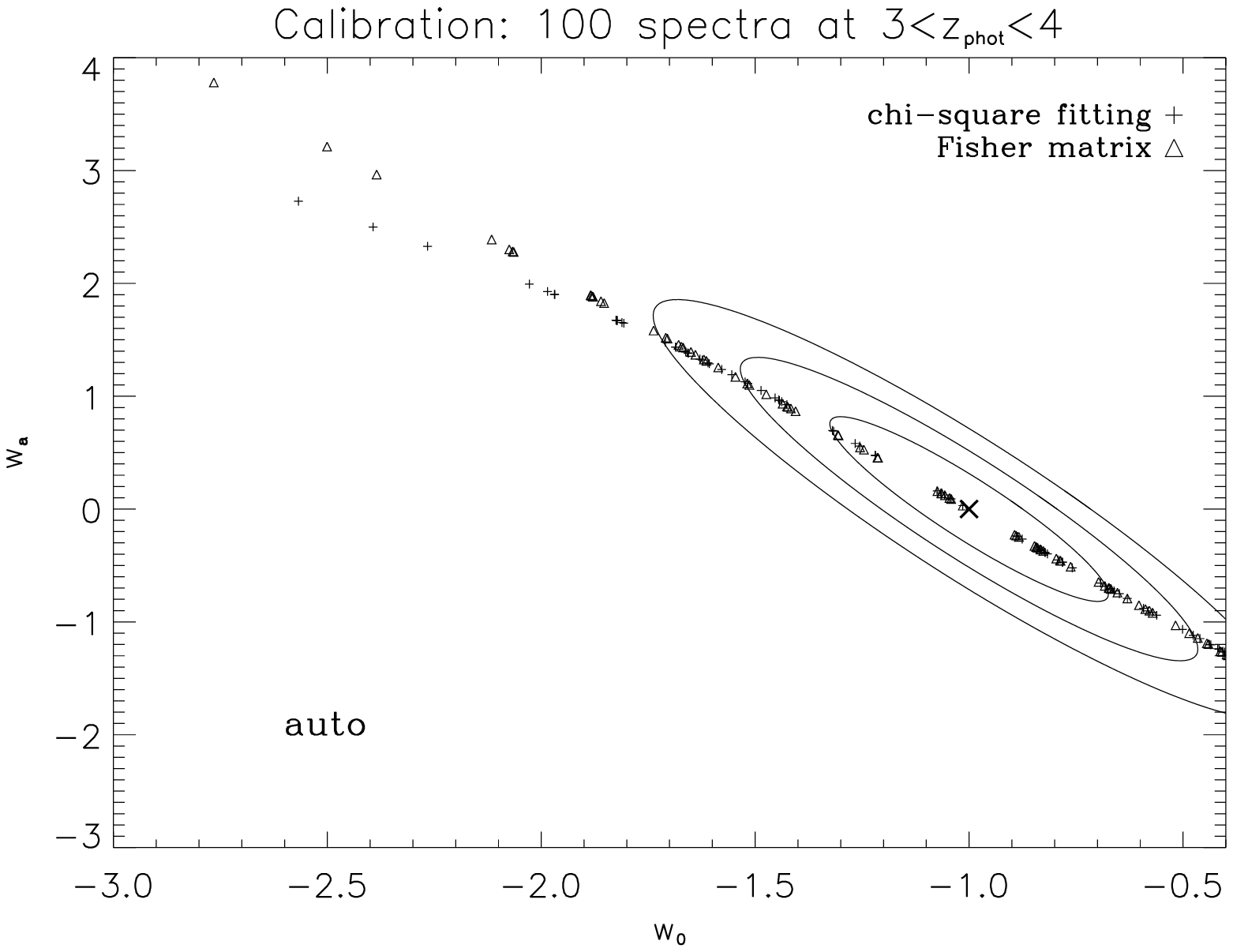}
\plottwo{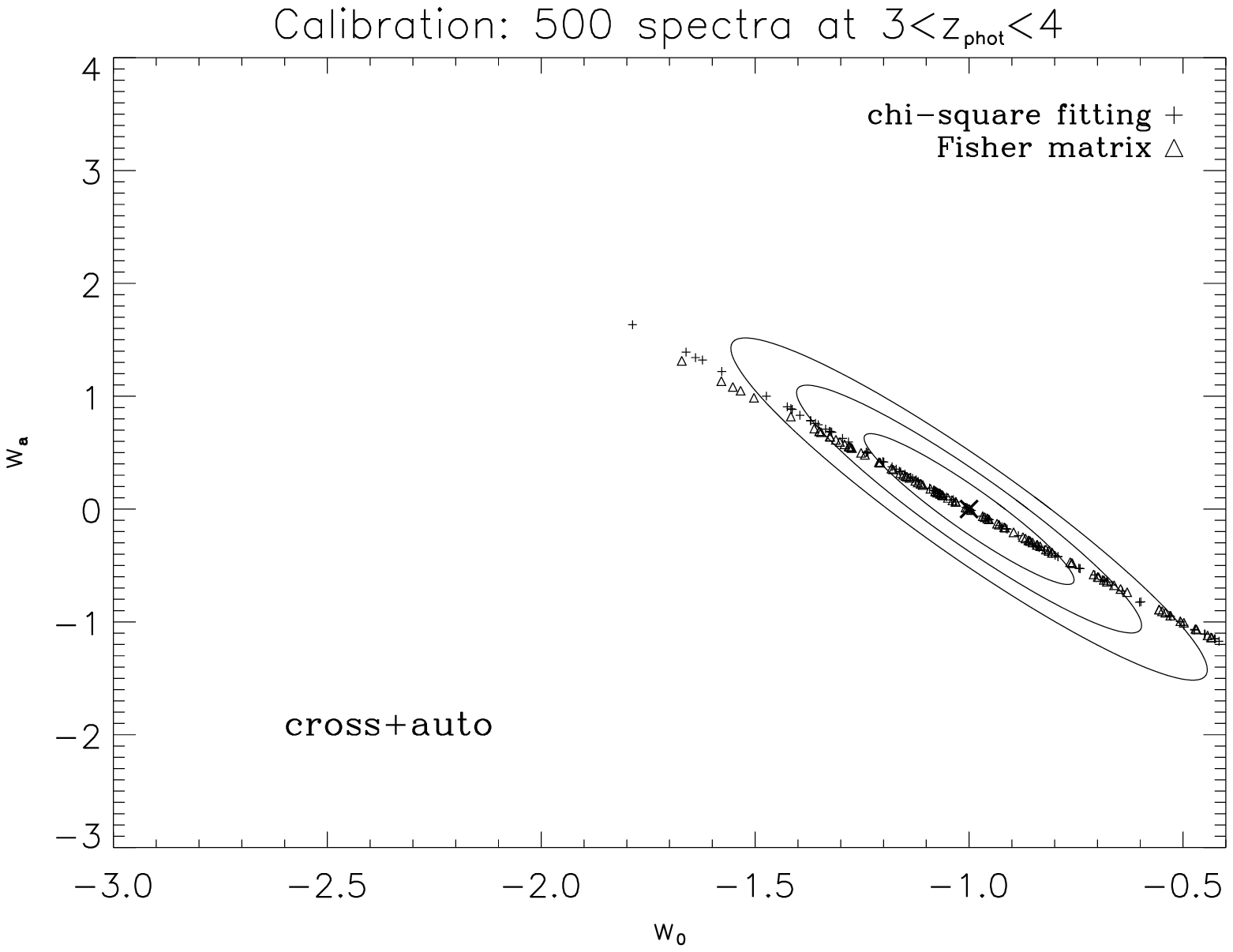} {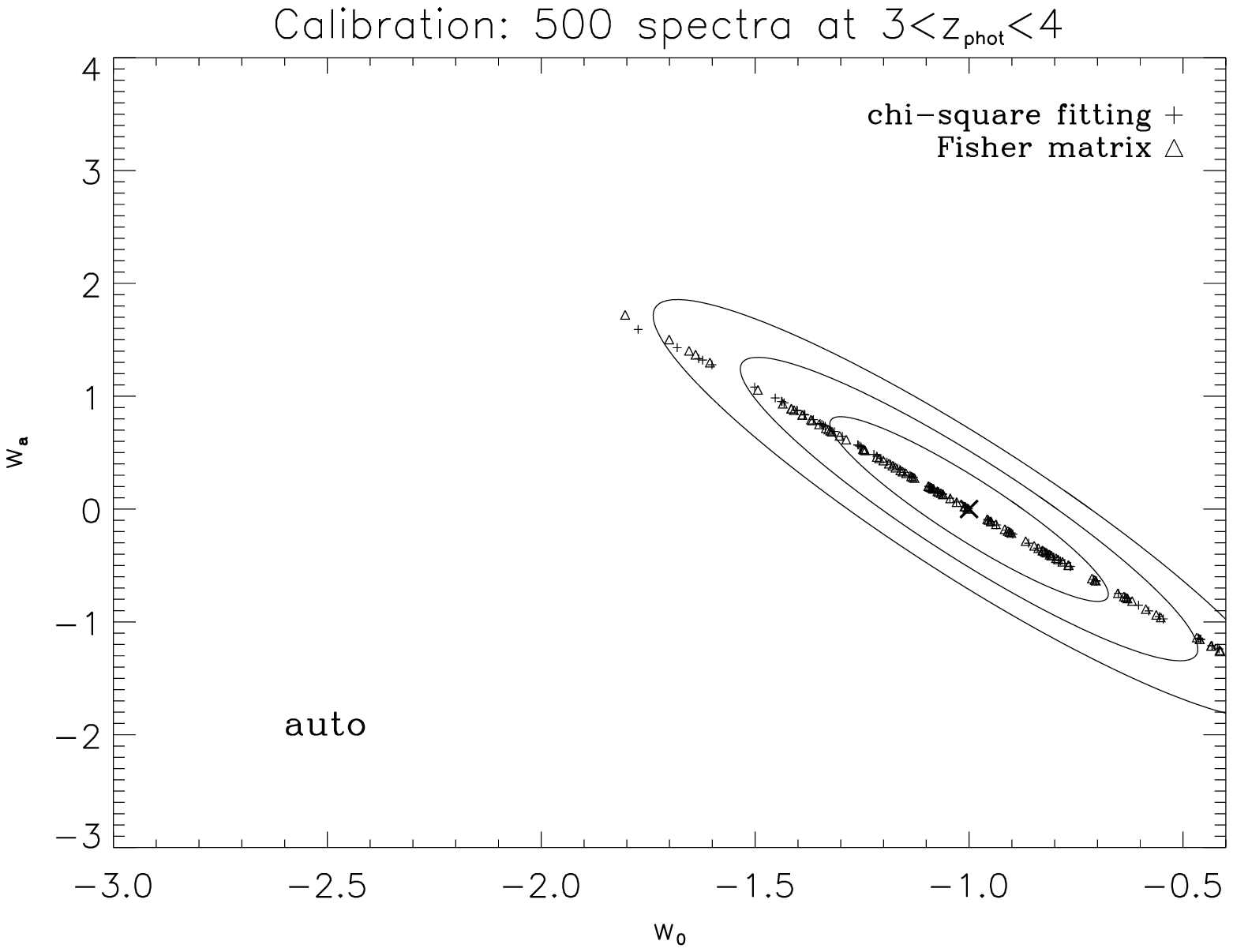}
\caption{\small Biases on $w_0$ and $w_a$
from $100$ realizations for $N_{\rm spec}=100$ (upper panels) and $N_{\rm spec}=500$ (lower panels),
respectively.
Left panels: the results of full calculations including
both the auto- and cross- correlations between different redshift bins. Right panels: the results considering only the auto-correlations within individual
redshift bins. In each panel, the pluses and triangles denote the `best fit' values of individual realizations from $\chi^2$ fitting and from the Fisher matrix calculations, respectively. The contours correspond to $1,2,3\sigma$ statistical errors computed at the fiducial values.}
\label{Fig:compare}
\end{figure}

Figure \ref{Fig:compare} shows the results on $w_0$ and $w_a$
of $100$ realizations for $N_{\rm spec}=100$ (upper panels) and $N_{\rm spec}=500$ (lower panels),
respectively. The symbols are the `best fit' values from individual realizations, and
the statistical errors are calculated at the fiducial values for the
purpose of clarity. The left panels are for the results of full calculations including
both the auto- and cross- correlations between different redshift bins. The
right panels are the results considering only the auto-correlations within individual
redshift bins. In each panel, the pluses and triangles correspond, respectively,
to the results from $\chi^2$ fitting and from the Fisher matrix calculations of Eq. (\ref{eq:bias}).
First when the bias is large and the fitted values are outside the $3\sigma$ contour,
the results from the linear approximation of Eq. (\ref{eq:bias}) start to deviate from the
$\chi^2$ fitting results, as seen most evidently in the upper right panel. This
shows the limitation of the linear approximation. If in some cases, very large biases are expected,
it is necessary to perform a full $\chi^2$ fitting to evaluate quantitatively the biases.
Otherwise, an over-estimate of the bias on $w_0$ and $w_a$ would be obtained from
the linear analysis. This is in good accordance with the results shown in Figure \ref{Fig:bias}.
On the other hand, when the biases are within the $3\sigma$ statistical contour,
the two sets of results are in excellent agreement. Thus in this regime,
the linear approximation of Eq. (\ref{eq:bias}) provides us an accurate and convenient way to
estimate the bias effects. We further find that the triangles from the linear calculations
are almost perfectly on a straight line with little scatters. The results from the
full $\chi^2$ calculations follow the same line when the biases are relatively small, and
are bent downward when the biases are large. Noting that the biases depend on
three quantities ($\bar{f}_{\rm cata}$, $\bar{z}_m$, $\bar{\sigma}$), the lining up of the
results indicate that there is a dominant factor among the three. Our detailed analyses,
presented later in Figure 6, show that the biases are mainly determined by
$(\bar{f}_{\rm cata}-{f}_{\rm cata})$, and
the shape of the redshift distribution in Bin-cata (described by the parameters
$\bar{z}_m$, $\bar{\sigma}$) has insignificant effects.

The straight line behavior presents us a quantitative way to describe the
joint bias of $w_0$ and $w_a$. We define the joint bias of $w_0$ and $w_a$ for a realization,
denoted by $\delta(w_0,w_a)$, as the distance in the ($w_0,w_a$) plane, between the fiducial point and the
best-fit point of this realization, with a sign equal to $sign[w_a(\rm fit)$ $-$ $w_a(\rm fiducial)]$. 
Furthermore, to quantify the relative bias with respect to the statistical errors,
we define the corresponding 
joint statistical error $\sigma(w_0,w_a)$ as the distance in the ($w_0,w_a$) plane between the fiducial point and the intersection point of the 
straight line with a given statistical error contour.
This definition accounts
correctly for the correlation between $w_0$ and $w_a$.

\begin{figure}
\epsscale{0.5}
\plotone{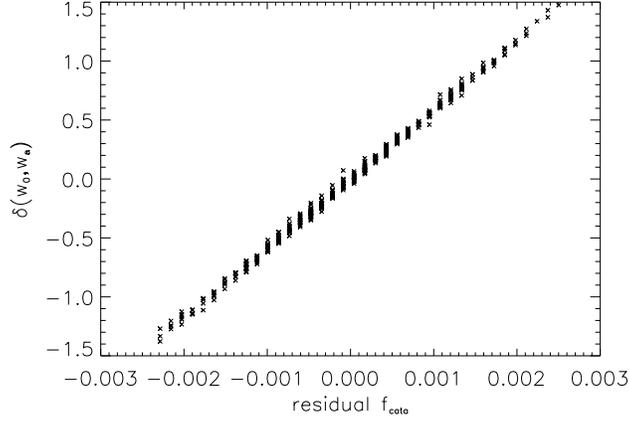}
\caption{ Joint $w_0$ and $w_a$ bias,
$\delta(w_0,w_a)$, and the residual $f_{\rm cata}$ (corresponding to
$f_{\rm cata}-\bar{f}_{\rm cata}$) from different realizations, for
$N_{\rm spec}=500$. The dispersion of
$\delta(w_0,w_a)$ at a fixed residual $f_{\rm cata}$ reflects the effect of different $\bar z_m$ and $\bar{\sigma}$ estimated from individual realizations.}
\label{Fig:real}
\end{figure}

To see the dependence of the bias on calibrations, in Figure \ref{Fig:real},
we show the correlation of the joint bias $\delta (w_0,w_a)$ defined above and the
residual $f_{\rm cata}$ (corresponding to $f_{\rm cata}-\bar{f}_{\rm
cata}$) from each individual realization for $N_{\rm spec}=500$. It clearly demonstrates that
the bias scales approximately linearly with the fraction of
catastrophic failures. The small dispersions of $\delta(w_0,w_a)$ at a
fixed residual $f_{\rm cata}$ correspond to results with
different calibrated $\bar{z}_m$ and $\bar{\sigma}$ for Bin-cata.
This reveals quantitatively that the effect of the specific
distribution within Bin-cata is sub-dominant, compared to the impact
of the overall fraction of Bin-cata, namely $\bar f_{\rm cata}$. This fact
actually allows us to obtain an approximate scaling relation for the dependence of the bias
on the survey area $f_{\rm sky}$ and on the catastrophic fraction $f_{\rm cata}$, which will
be discussed shortly.

\begin{figure}
\epsscale{0.5}
\plotone{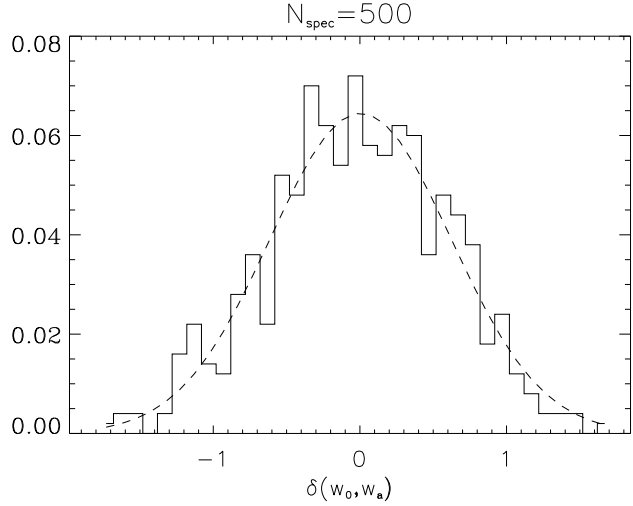}
\caption{ The distribution of joint $w_0$ and $w_a$ bias $\delta(w_0,w_a)$ derived from 500 realizations of calibration sampling with $N_{\rm spec}=500$. The dashed line shows a best-fit Gaussian curve to the bias distribution histogram.}
\label{Fig:rms}
\end{figure}

\begin{figure}
\epsscale{0.5}
\plotone{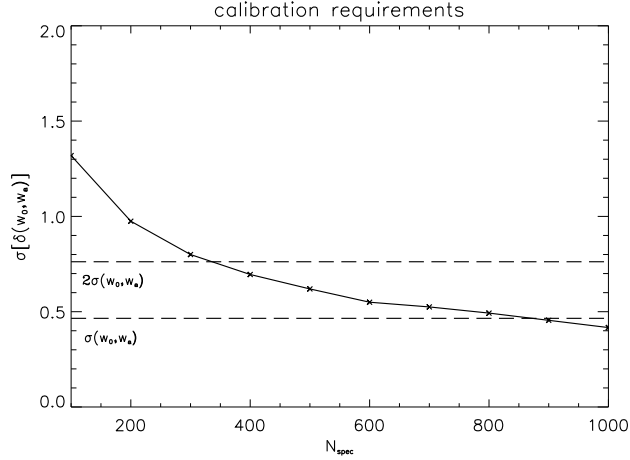}
\caption{Relation of the rms bias,
$\sigma[\delta(w_0,w_a)]$,  with the number of spectra at $3\le z_{\rm phot}\le 4$
available for calibration, $N_{\rm spec}$. The dashed lines denote the joint
$1\sigma$ and $2\sigma$ statistical errors of $w_0$ and $w_a$, respectively.}
\label{Fig:scaling}
\end{figure}

With the biases obtained from individual realizations, we can
statistically quantify the typical level of bias for a given $N_{\rm spec}$.
From Figure \ref{Fig:compare} and \ref{Fig:real}, we see
that the average bias from all the realizations of a given $N_{\rm spec}$
is expected to be close to zero. Hence we use the root-mean-square bias calculated 
from the bias distribution to characterize the bias level for that particular $N_{\rm spec}$.
To illustrate it more clearly, in Figure \ref{Fig:rms}, we show the distribution of 
$\delta(w_0,w_a)$ from $500$ realizations for $N_{\rm spec}=500$. 
The bias for each individual realization is calculated
by employing the Fisher matrix formalism Eq. (\ref{eq:bias}) since it can give good approximation
when the residual $f_{\rm cata}$ is reasonably small. It can be seen that 
the distribution is very close to a zero-mean Gaussian distribution. The rms bias
is calculated to be $\sigma[\delta(w_0,w_a)]\approx 0.61$ for $N_{\rm spec}=500$.
We perform such a calculation on $\sigma[\delta(w_0,w_a)]$ for each considered $N_{\rm spec}$,
and the results are shown in Figure \ref{Fig:scaling}.
We can see that $\sigma[\delta(w_0,w_a)]\propto 1/\sqrt{N_{\rm spec}}$ 
to a very good approximation, and can be written as 
$\sigma[\delta(w_0,w_a)]=0.61/\sqrt{N_{\rm spec}/500}$.
For $f_{\rm sky}=0.025$ and $f_{\rm cata}=0.01$, the $1\sigma$ and $2\sigma$
$w_0$ and $w_a$ joint statistical errors defined above are $\sim 0.465$ and 
$0.762$, respectively. Then the needed $N_{\rm spec}> 860$ and $320$
so that the rms bias $\sigma[\delta(w_0,w_a)]$
can be smaller than the $1\sigma$ and $2\sigma$ statistical errors, respectively.

The specific results on the required $N_{\rm spec}$ shown above are calculated
with $f_{\rm sky}=0.025$ and $f_{\rm cata}=1\%$. However, based on our analyses, we
can derive a scaling relation which permits us to estimate the needed $N_{\rm spec}$
for different $f_{\rm sky}$ and $f_{\rm cata}$.
As illustrated in Figure \ref{Fig:real}, the bias is linearly proportional to
$\bar f_{\rm cata}-f_{\rm cata}$. Thus the rms bias
$\sigma[\delta(w_0,w_a)]$ should be approximately proportional to
the rms of $\bar f_{\rm cata}-f_{\rm cata}$, namely $\sigma (\bar f_{\rm cata}-f_{\rm cata})$.
Furthermore, for a set of realizations, the number of calibrating galaxies that are
identified to be in Bin-cata follows the binomial distribution with a mean value of 
$N_{\rm spec}p$ and the variance of $N_{\rm spec}p(1-p)$, where 
$p=f_{\rm cata}\times N_{\rm tot}/\Delta N_5$,
with $N_{\rm tot}$ and $\Delta N_5$
representing the total number of galaxies and the number of galaxies
with $3\le z_{\rm phot}\le 4$, respectively. 
In our consideration, we have $p\sim 0.15<<1$, and the distribution is close to 
the Poisson distribution. Then 
$\sigma (\bar f_{\rm cata}-f_{\rm cata})\propto \sqrt{f_{\rm cata}}/ \sqrt{N_{\rm
spec}}$. On the other hand, as we discussed in \S 3, the absolute bias should not
depend on the survey area. The statistical error $\sigma(w_0,w_a)$ scales as $1/\sqrt {f_{\rm sky}}$.
Therefore the relative bias with respect to the statistical error scales as
$\sqrt {f_{\rm cata}\times f_{\rm sky}/ N_{\rm spec}}$. This in turn gives rise to the
scaling relation for the needed $N_{\rm spec}\propto f_{\rm cata}\times f_{\rm sky}$.
We then have
\begin{equation}
N_{\rm spec}  > 860\left ({f_{\rm cata} \over 0.01} \right ) \left ({f_{\rm sky} \over 0.025} \right )
\label{eq:nspec}
\end{equation}
for keeping the bias smaller than the $1\sigma$ statistical error. Thus
for $f_{\rm cata}=1\%$ and $f_{\rm sky}=0.25$ ($A=10,000\hbox{ deg}^2$), 
we would need $N_{\rm spec}  > 8600$. 
We emphasize that the number $860$ in the scaling relation above, depends on the specific detail of the considered  catastrophic errors  and the overall galaxy redshift distribution $n(z)$ 
[see Eq. (\ref{eq:bias})].  
Our analyses are particularly for the island located at $(z_{\rm
spec},z_{\rm phot})\sim(0.4,3.5)$, which is the dominant
catastrophic error occurrence seen from our photo-z simulations. 
It is known that the pattern of catastrophic errors depends on specific 
photo-z measurements, notably on the wavelength coverage and the filter set used.  
Thus different sets of filters would result in different catastrophic errors, 
and consequently different calibration requirements. 
It is noted that the above linear scaling relation for $f_{\rm cata}$
is an approximate one as the normalization ($860$ here) has a mild dependence on
the fiducial value of $f_{\rm cata}$. It should also be pointed out that the
calibration requirement we present does not contain the part needed
internally for the photo-z measurements, which we assume have been done 
with considerable accuracy, prior to yielding the $f_{\rm cata}=1\%$ 
catastrophic outliers.

\section{Summary and Discussion}
\label{sect:dis}

In this paper, we concentrate on the island part of the galaxies in
the $(z_{\rm spec},z_{\rm phot})$ plot where $z_{\rm phot} $ is
completely mis-estimated. This type of large catastrophic failures
in the photometric redshift measurements arise from the
confusion of different characteristics of galaxies' SED due to the
limited wavelength coverage. The most notable features used in the
photo-z measurement of galaxies are the Lyman and $4000 \AA$ breaks.
The mis-identification of the two results in the islands
at $(z_{\rm spec}, z_{\rm phot})\sim (0.4, 3.5)$ and $(z_{\rm spec},
z_{\rm phot})\sim (3, 0.2)$ seen in Figure 1.
We particularly analyze the systematic bias induced by such islands on cosmological parameter
estimations. We find that a fraction of $f_{\rm cata}=1 \%$ catastrophic failures
due to galaxies at $z_{\rm spec}\sim 0.4$ being misidentified to be
at $z_{\rm phot}\sim3.5$, which is the dominant case of catastrophic
failure occurrence in our photo-z simulations, can bring systematic
biases on $w_0$ and $w_a$ at a level far exceeding the statistical
errors in $5$-tomographic-bin analysis.

It has been realized that spectroscopic redshift calibration may
be necessary for future weak lensing surveys in order to reduce the
uncertainty of photometric redshift errors. The large bias shown in
our analyses reinforces such an argument. We investigate the
requirements of the spectroscopic calibration to retrieve the
unbiased parameter estimates. We find that for $f_{\rm cata}=1\%$,
and $f_{\rm sky}=0.025$, about 400-1000 spectral redshift
measurements for galaxies with $z_{\rm phot}=[3,\ 4]$ are needed. We
further present a scaling relation of $N_{\rm spec}\propto {f_{\rm
cata}\times f_{\rm sky}}$. Thus for a future survey with a survey
area of $10,000 \hbox{ deg}^2$, the size of the calibration sample
in the redshift range $z_{\rm phot}=[3,\ 4]$ should be on the order
of $10^4$ if $f_{\rm cata}$ remains to be $\sim 1\%$. In analyzing
the error degradation from general photometric errors, people have
argued that the inaccuracy of the high redshift photometric
measurements can be compensated by increasing the measurement
accuracy for galaxies at relatively low redshifts. Therefore much
less spectroscopic calibration samples at high redshifts are needed
(e.g., Ma et al. 2006). Concerning the catastrophic-error-induced
bias discussed in this paper, however, there is no such a
compensation. Thus the required $\sim 10^4$ spectral calibrations in
$z_{\rm phot}=[3,\ 4]$ may be regarded as challenging tasks.

It is also argued that the inclusion of both U and near-IR filters
is essential for the accuracy of photo-z and the minimization of the
catastrophic outliers (e.g., Dahlen et al. 2008, Abdalla et al.
2008), as the Lyman and $4000 \AA$ breaks can be effectively
distinguished. Given that our simulations with $9$-filter set
already include the near-IR bands, we consider the improvements from
U band coverage. Extending the wavelength coverage to $3200 \AA$,
our simulations show that the catastrophic failure fraction
decreases dramatically to the level around $0.1\%$ (Jouvel et al.
2009, in prep.). In this case, the bias should no longer be
significant compared to the statistical errors. It is thus important
to investigate the possibility to include $<3500 \AA$ sensitivity
for the next generation space dark energy mission.

\begin{figure}
\epsscale{1}
\centering
\plottwo{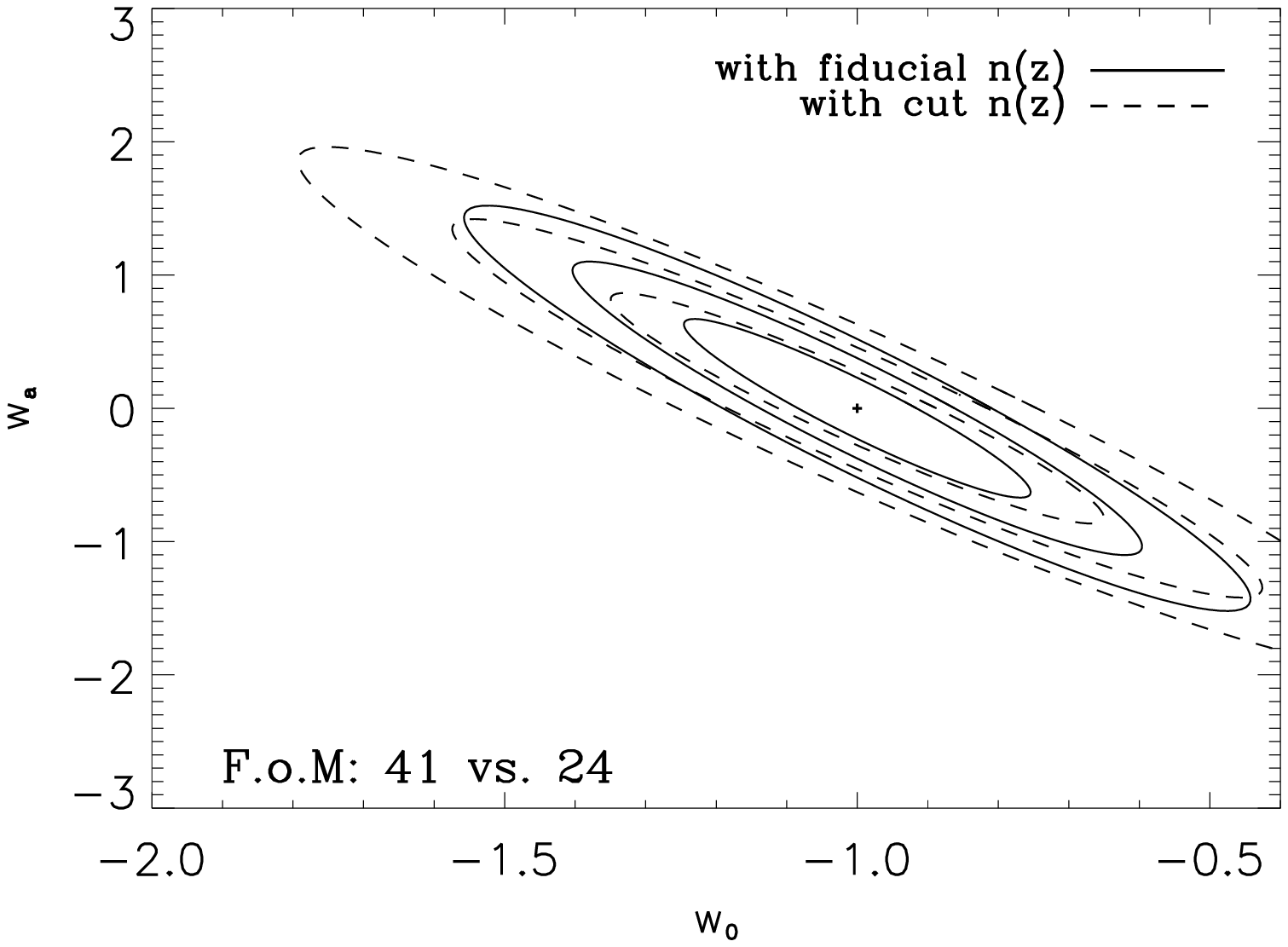}{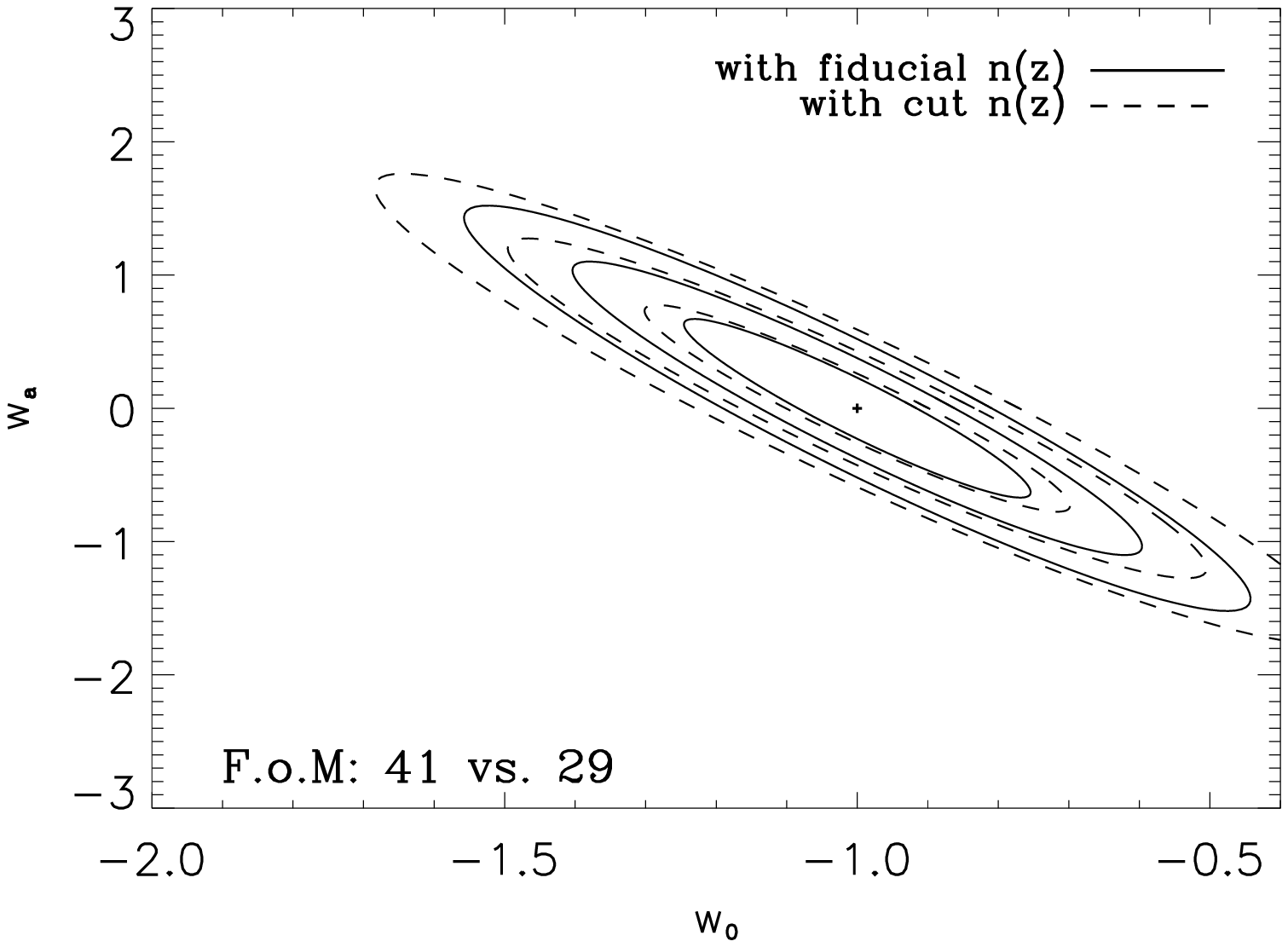}
\caption{The
comparisons of constraints on $w_0$ and $w_a$. The solid contours in
both panels are the results for the fiducial 5 bin tomography. The
dashed contours are the results for discarding those galaxies at
$z<0.5$ and $z>2.5$, either with the left 3 bins at z=[0.5, 2.5]
kept unchanged (in left panel) or with galaxies at z=[0.5, 2.5]
re-divided into $5$ finer bins (in right panel). Gaussian priors
$\sigma(\Omega_b)=0.01$ for $\Omega_b$ and $\sigma(p_i)=0.05$ are
applied upon all other hidden parameters.}
\label{Fig:cut}
\end{figure}

It has also been proposed to filter out those problematic galaxies,
such as late type galaxies, low signal-to-noise galaxies, or very
low/high redshift (e.g., $z<0.5$/$z>2.5$) galaxies whose photo-zs
are most suspicious, in the shear analysis to reduce the impact of
catastrophic failures (Abdalla et al. 2008; Margoniner \& Wittman
2008; Heavens, Kitching \& Taylor 2006). In such cases, the
statistical power of a survey is inevitably lowered to certain
extents depending on the fraction of galaxies discarded. In Figure
\ref{Fig:cut}, we show the degradation of the constraints if those
galaxies at $z<0.5$ and $z>2.5$ (thus 3 bins left) are excluded in
the analyses, compared to the case with calibrated (known) $f_{\rm
cata}$. We see that the Figure of Merit (denoted as
F.o.M)\footnote{F.o.M = $\sqrt {\rm det (F)}$, where F is the
reduced Fisher matrix for $w_0$ and $w_a$ after marginalization over
other parameters.} decreases from 41 to 24, i.e., almost a factor of
two degradation. If we re-divide the galaxies with $0.5\le z\le 2.5$
into $5$ finer bins (right panel of Figure \ref{Fig:cut}), the
F.o.M = 29, or a degradation factor $\sim 1.5$ (see also table 1 for a summary.). 
All these calculations are done with our fiducial survey conditions and Gaussian priors $\sigma(\Omega_b)=0.01$ for $\Omega_b$ and $\sigma(p_i)=0.05$ are applied upon all other parameters.

In our current analyses, we isolate the island galaxies appearing around
$(z_{\rm spec}$,$z_{\rm phot})=(0.4,3.5)$ to see clearly their
effects on cosmic shear analysis. On the other hand, the scatters
and small bias of $z_{\rm phot}$ around $z_{\rm spec}$ have been
modeled and studied in, for example, Ma et al. (2006). With
$\sigma_z$ and $z_{\rm bias}$ as well as their uncertainties
$\Delta\sigma_z$ and $\Delta z_{\rm bias}$ taken into account, the
statistical errors of the constrained cosmological parameters will
be degraded depending on the values of ($\sigma_z$, $z_{\rm bias}$,
$\Delta\sigma_z$, $\Delta z_{\rm bias}$). For instance, considering
($\sigma_z$, $z_{\rm bias}$, $\Delta\sigma_z$, $\Delta z_{\rm
bias}$) $=(0.05(1+z), 0, 0.003, 0.003)$, the degradation factor for
$w_0$ and $w_a$ is about $1.5$ (Ma et al. 2006). Then a crude
estimate based on our analyses gives the needed $N_{\rm spec}$ at
$z\sim[3,\ 4]$ to be about $8600/(1.5)^2\sim 4000$ to control the
bias down to $1\sigma$ level for $f_{\rm cata}=1\%$ and $f_{\rm
sky}=0.25$. More complete analyses should be done for both the
degradation and the bias simultaneously taking into account both the
small and the large photo-z errors.

\acknowledgements We thank the referee for the very constructive comments and suggestions,
which help to improve the paper greatly.
We are grateful to Dragan Huterer, Gary Bernstein, Anne Ealet, Alexie
Leauthaud, Adam Amara, Hu Zhan, Liping Fu, Vera Margoniner, Sarah
Bridle and Ludovic Van Waerbeke for their helpful discussions. We thank
Craig Markwardt for his released chi-square fitting software 'MPFIT'
and his very useful suggestions on some technical aspects. This
research is supported in part by the NSFC of China under grants
10373001, 10533010 and 10773001, and the 973 program
No.2007CB815401. LS also benefits from a scholarship from the
"Minist\`ere des Affaires Etrang\`eres " of the French government.
SJ and JPK acknowledge support from CNRS and CNES. CT, LS, and AT 
acknowledge support from CNRS.

\begin{table}
  \caption[]{Comparisons of F.o.M of $w_0$ and $w_a$. Note, the fiducial F.o.M is given under a calibrated (known) $f_{\rm cata}$. Gaussian priors $\sigma(\Omega_b)=0.01$ for $\Omega_b$ and $\sigma(p_i)=0.05$ are applied upon all other parameters.}
  \label{Tab:cata}
  \begin{center}\begin{tabular}{clcl}
  \hline\noalign{\smallskip}
        &  Fiducial 5-bins   &\ \ \ \ \ \ \ \ \ \ \ \ \ \ \ \ \ \ \ \ \ \ $\mid$\ \ \  With cut and $0.5<z<2.5$    \\
  \hline\noalign{\smallskip}                
        &     & $\mid$ \ \ 3-bins left & $\mid$ \ \ 5-bins re-defined \\ 
  \hline\noalign{\smallskip}               
  F.o.M & \ \ \ \ 41    &  \ \ \ \ \ \ \ \ 24    & \ \ \ \ \ \ \ \ \ 29          \\
  \noalign{\smallskip}\hline 
  \end{tabular}\end{center}
\end{table}

\end{document}